\documentclass[jcp,twocolumn]{revtex4}
\usepackage[utf8x]{inputenc}
\usepackage{amssymb}
\usepackage{amsmath}
\usepackage[dvips]{graphicx}
\usepackage{bm}

\newcommand{\beq}{\begin{equation}}
\newcommand{\eeq}{\end{equation}}
\newcommand{\ba}{\begin{eqnarray}}
\newcommand{\ea}{\end{eqnarray}}
\newcommand{\bc}{\begin{center}}
\newcommand{\ec}{\end{center}}

\newcommand{\m}{\mbox{\boldmath ${\mu}$}}
\newcommand{\oOmega}{\mbox{\boldmath ${\Omega}$}}
\newcommand{\tm}{\tilde}

\begin{document}
\title{Dynamics of fibers in a wide microchannel}
\author{Agnieszka M. S\l owicka}
\author{Maria L. Ekiel-Je\.zewska}
\author{Krzysztof Sadlej}
\author{Eligiusz Wajnryb}

\date{\today}

\address{%
Institute of Fundamental Technological Research,
Polish Academy of Sciences,
Pawi\'nskiego 5B,
02-106, Warsaw,
Poland\\
}%

\begin{abstract}
Dynamics of single flexible non-Brownian fibers, tumbling in a Poiseuille flow between two parallel solid plane walls, is studied with the use of the {\sc hydromultipole} numerical code, based on the multipole expansion of the Stokes equations, corrected for lubrication. It is shown that for a wide range of the system parameters, the migration rate towards the middle plane of the channel  increases for  fibers, which are closer to a wall, or are more flexible (less stiff), or are longer.  The faster motion towards the channel center is accompanied by a slower translation along the flow and a larger fiber deformation.
\end{abstract}
\maketitle

\section{Introduction}
Flexible fibers dynamics has been the subject of interest in a myriad of problems ranging from industrial applications around the paper making process or water purification down to the level of single DNA molecules \cite{Smith1999,Du2005}. There are articles, which concentrate on the behavior of suspensions and interaction between fibers leading to grouping and entanglement and the flocculation phenomena \cite{Klingenberg2004,Guazzelli1999}.
Others look closer at physical mechanisms of lateral, cross-stream migration, fiber dynamics and stretching in the flow, due to shear or wall effects \cite{Hagerman1981,Yamamoto1993,Yamamoto1995,Klingenberg1997_107,Tornberg2004,Schroeder2005}.

The single fiber problem has attracted a lot of attention in recent years due to relevance to many new experimental results conducted in microfluidic devices and implications on DNA dynamics and its biological role on gene regulation \cite{Cloutier2004}. DNA flexibility is extremely important for its functioning. Looping of the fibers allows proteins bounded at distant sites to act synergistically \cite{Du2005}. An insight into these important functions is now provided by experimental techniques which have evolved to a point enabling one to directly visualize and manipulate the conformational changes imparted by a fluid flow on single, flexible DNA polymers \cite{Teixeira2005,vanMameren2008}. Such experiments have shown the conformational dynamics of individual, flexible polymers in steady shear usually exhibit temporal fluctuations described as end-over-end tumbling of the molecule \cite{Smith1999}.

Many studies have also concentrated on the phenomenon of cross-stream migration in confined flows due to its relevance in nano- and microfluidic devices and fundamental practical significance \cite{Chen2005,Winkler2006,WinklerEPL2008,WinklerEPL2010,LaddLB2010, Ladd2005}. Lateral migration, which in low-Reynolds-number dynamics is absent for rigid particles, is found in the evolution of flowing flexible fibers due to dynamical changes of shape while translating along the channel. It is accepted that confinement has a crucial impact on the behavior of single polymers \cite{WinklerEPL2010,Zurita} by effects of spatially varying shear rate and hydrodynamic interactions with the impenetrable walls. In general, the flow properties of flexible fibers have been shown to be highly influenced by hydrodynamic interaction, both within the fiber and with the system boundaries \cite{Ladd2006}.

Although single-particle experiments have provided a leap in the visualization of dynamics of fibers in flows, in particular the dynamics of a DNA molecule or a nanofiber\cite{Sadlej-Kowalewski}, theoretical concepts and numerical models are often still the key to understanding the underlying physical mechanisms.  The so called bead model \cite{DhontBOOK} has now been accepted to reproduce the hydrodynamic properties of a flexible  fiber. Each pair of bonded spheres can stretch and bend, by changing bond distance and bond
angle, respectively. The strength of bonding, or flexibility of the fiber model, is defined typically by
two parameters of stretching and bending. Altering these parameters, the property of the fiber model can be
changed from fully rigid to flexible. It is used widely  to model flexible fibers in various simulation techniques ranging from Monte
Carlo simulations \cite{Podtelezhnikov2000}  to Brownian \cite{Liu1989, Symeonidis2005} and molecular dynamics
\cite{Gompper2010}.

In this paper, we describe a single fiber with the use of the bead model, and the {\sc{hydromultipole}} numerical code \cite{MEJ-EW}, which implements the theoretical multipole method \cite{Cichocki2000,Ekiel-Wajnryb-book} of calculating hydrodynamic interactions between the particles in Stokes flows. We assume that the Brownian motion of the fiber segments is negligible, and approximate
the channel boundaries by two infinite parallel solid walls. 
We evaluate hydrodynamic interactions between the segments of a flexible fiber, and apply the single-wall superposition \cite{Blawzdziewicz-Wajnryb} to take into account the influence of the channel boundaries, 
 following the theoretical approach and the numerical procedure outlined in Ref.~\cite{Sadlej-Kowalewski}. 

Here we focus on the guiding question how to increase the migration rate of the fiber towards the middle plane of the channel, by an adequate choice of the system parameters. In Sec.~\ref{II}, we describe the system and explain our model. In Sec.~\ref{III}, we present the results for the fiber motion and its shape evolution. 
The main conclusions are outlined in Sec.~\ref{IV}.

\section{Motion of fibers in Poiseuille flow: theoretical model}\label{II}

\subsection{Problem}

In the paper, we consider a single mobile and flexible microfiber immersed in a fluid flow. We assume that  velocity of the flow is very small,  with the Reynolds number Re $\ll$ 1, small enough to neglect
inertia effects. Here Re is the product of a particle velocity and its size, divided by the fluid kinematic viscosity. We assume that the Peclet number is large, Pe $\gg$ 1, and Brownian 
motions are irrelevant.  Here Pe is the product of a fiber velocity and its size, divided by the diffusion constant. We assume also the absence of unsteady effects. For such a system, the hydrodynamic friction 
forces and torques exerted by the moving fibers on the fluid are equal to the external forces and torques imposed on them.

For the system specified above, the fluid velocity $\mathbf{v}$ and pressure $p$ satisfy the stationary Stokes equations \cite{KimKarrila,Happel},
\ba
\eta {\bm\nabla }^{2}\mathbf{v-\bm\nabla }p &=&{\bf 0},  \label{001} \\
\mathbf{\bm\nabla \cdot v} &=&0,  \label{001a}
\ea
where $\eta$ is the fluid dynamic shear viscosity. 

\begin{figure}[ht]
\includegraphics[width=6.7cm]{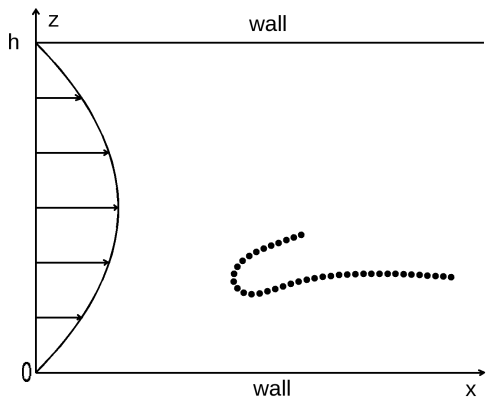}
\caption{The system.}\label{fig:fiber}
\end{figure}

The fluid is confined inside a microchannel made of two parallel infinite solid walls at $z=0$ and $z=h$, as shown on Fig. \ref{fig:fiber} The walls are non-deformable.
\\
The external fluid flow $\mathbf{v}_0$ inside the channel is a Poiseuille flow,
\beq
{\bf v}_0= 4  v_{m} z(h-z)/h^2\,\hat{\bf x}.
\label{eq:Poiseuille}
\eeq
Note that ${v}_m$ is the the maximum fluid velocity, attained at $z=h/2$. The above set of partial differential equations is supplemented by the stick boundary conditions at the
surface of the fiber and at the hard walls, which confine the fluid. At infinity, we assume that  $\mathbf{v}-\mathbf{v}_0={\bf 0}.$ 

In this paper, the fiber is initially straight and aligned with the flow, and we detect its motion and shape deformation. To this goal, we first need to specify the elastic and 
bending forces which keep together all the fiber segments. These are analyzed in the succeeding subsection.

\subsection{Model of a fiber: elastic and bending forces}
\label{model}
To determine the fiber motion, the bead model is used~\cite{DhontBOOK}. Each fiber strand is constructed out of $N$  solid non-deformable spherical particles of diameter $d$ which can move with respect to each other. 
The relative motion of the beads is constrained by, and results from elastic and bending forces, which are discretized as described in Ref.~\cite{Stark2006}. 

\begin{figure}[ht]
\includegraphics[width=6.2cm]{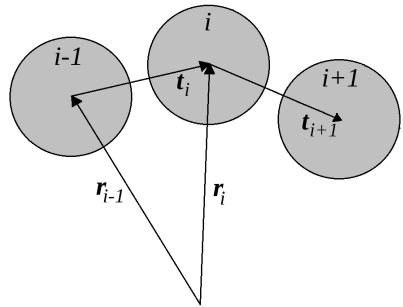}
\caption{Bead-spring model of a fiber~\cite{Stark2006}. Each bead feels non-hydrodynamic forces due to the stretching and bending.}\label{fig:fig}
\end{figure}

The elastic (extension) energy between all the neighboring beads  is given by the Hook's law,
\beq
  E^{e} = \frac{1}{2}\tm{k}\sum_{i=2}^{N} \left(l_i-\tm{l}_0\right)^2,
\eeq
where $\tm{l}_0$ is the equilibrium distance between beads, $\tm{k}$ is the Hooke's constant and 
\beq
l_i = |\bm{t}_i|, 
\eeq
where
\beq
 \bm{t}_i = \bm{r}_i- \bm{r}_{i-1},
\eeq
and $\bm{r}_i$ is the position of the bead $i$ center (see Fig. \ref{fig:fig}).

The extensional force exerted on the bead $i$ is then equal minus the gradient of the elastic energy,
\beq
  \bm{F}^e_i = -\bm\nabla_i E^{e} = -\tm{k}(l_i-\tm{l}_0)\hat{\bm{t}}_i + \tm{k} (l_{i+1}-\tm{l}_0)\hat{\bm{t}}_{i+1},
\eeq
where $\bm\nabla_i$ is the derivative with respect to $\bm{r}_i$ and $\hat{\bm{t}}_i = \bm{t}_i/l_i$.

The bending energy in case of a continuous model of a fiber is given by the integral, \cite{Landau}
\beq
 E^{b} = \frac{1}{2}{\tm{A}}\int_{0}^{L} \left(\frac{d\hat{\bm{t}}}{ds}\right)^2 ds,
 \label{eq:bend_energy_cont}
\eeq
where $\tm{A}$ is the bending stiffness.

The discretized version of Eq. (\ref{eq:bend_energy_cont}), describing the bead model \cite{Stark2006}, is 
\beq
 E^{b} = \frac{1}{2}\tm{A}\sum_{i=1}^{N} f_i \frac{\left(\hat{\bm{t}}_{i+1} - \hat{\bm{t}}_{i}\right)^2}{\tm{l}_0} =
        \frac{\tm{A}}{\tm{l}_0} \sum_{i=1}^{N} f_i \left(1-\hat{\bm{t}}_{i+1} \cdot \hat{\bm{t}}_{i}\right),
\eeq
where
\beq
  f_i = \left\{ \begin{array}{ccc}
                 1 & \textrm{for} & 2\leq i\leq N-1,
                 \\
                 0 & \textrm{for} & 1,N.
                 \end{array}\right. 
\eeq

The corresponding bending force $\bm{F}^b_i = -\bm\nabla_i E^{b}$ is then equal to 
\hspace{-0.9cm} \ba \hspace{-0.9cm}
\!\!\! &&\!\!\! \!\!\!\!\!\!\bm{F}^b_i = \frac{\tm{A}}{\tm{l}_0}\Bigg\{\!
                 \frac{f_{i-1}}{l_i}\hat{\bm{t}}_{i-1}-\!\left[ \frac{f_{i-1}}{l_i}\hat{\bm{t}}_{i-1} \cdot \hat{\bm{t}}_{i}+\frac{f_{i}}{l_{i+1}}
                                                             +\frac{f_{i}}{l_i}\hat{\bm{t}}_{i}\cdot \hat{\bm{t}}_{i+1} \right]\!\hat{\bm{t}}_{i}
                                                             \nonumber\\
                                                     \hspace{-0.9cm}  \!\!\!  &&\!\!\!\!\!\!\!\!\!
                                                       +\!\left[ \frac{f_{i}}{l_{i+1}}\hat{\bm{t}}_{i}\cdot \hat{\bm{t}}_{i+1}+\frac{f_{i}}{l_{i}}
                                                             +\frac{f_{i+1}}{l_{i+1}}\hat{\bm{t}}_{i+1} \cdot\hat{\bm{t}}_{i+2} \right]\!\hat{\bm{t}}_{i+1} -      \frac{f_{i+1}}{l_{i+1}}\hat{\bm{t}}_{i+2}
\! \Bigg\}.
\nonumber\\
\ea

Summarizing, the total non-hydrodynamic force exerted on each bead $i = 1,...,N$ consists of the extensional force and the bending force,
\beq
 \bm{F}_i = \bm{F}^{e}_i+\bm{F}^b_i.\label{Fi}
\eeq
Note that the total non-hydrodynamic 
force applied to all the beads of the fiber is zero,
\beq
  \sum_{i=1}^{N}\bm{F}_i = 0.
  \label{eq:F_zero}
\eeq

From now on we use dimensionless variables. 
Distances are normalized by the bead diameter $d$, velocities by the maximal velocity $v_m$ of the ambient Poiseuille flow, and the forces by 
\beq
f_0=\pi\eta d v_m.\label{funit}
\eeq
 Time is given in units of  
\beq
t_0 = d/v_m.\label{tunit}
\eeq
We introduce dimensionless parameters:
\ba
 k&=&\tm{k}d/f_0,\label{kdef}\\
 l_0&=& \tm{l}_0/d,\\
 A&=& \tm A/(f_0d^2)\label{Adef}.
\ea
The parameter $A$ describes how large is the bending energy (or bending force) of the fiber, if the external fluid flow \eqref{eq:Poiseuille} is fixed. A larger value of $A$ means a stiffer  (less flexible) fiber, and this is why A is called a bending stiffness parameter. At $A=0$ the fiber may bend freely without the use of energy. 
In fact, $A$ is the ratio of the bending force to a hydrodynamic force related to the ambient flow amplitude $v_m$. In a similar way, $k$ is the ratio of the elastic force to a hydrodynamic force related to the ambient flow amplitude $v_m$.

\subsection{Fiber dynamics: theoretical and numerical methods}
Assume first that the fiber is fixed at a given configuration, and that the forces ${\bf F}_0=({\bf F}_{01},...,{\bf F}_{0N})$ and torques ${\bf T}_0=({\bf T}_{01},...,{\bf T}_{0N})$ exerted by the flow on each bead are known. 
For a mobile fiber,  ${\bf F}_0$ and ${\bf T}_0$ are used to determine the translational and rotational velocities of all the fiber beads, $i=1,...,N$, 
$\mathbf{U}=(\mathbf{U}_1,...,\mathbf{U}_N)$ and $\mathbf{\Omega}=(\mathbf{\Omega}_{1},
...,\mathbf{\Omega}_{N})$, respectively. $\mathbf{U}$ and $\mathbf{\Omega}$  depend linearly on the elastic and bending forces, ${\bf F}=({\bf F}_1,...,{\bf F}_N)$, with ${\bf F}_i$ given by Eq.~\eqref{Fi}, and on ${\bf F}_0$ and ${\bf T}_0$,
\beq
\left( \begin{array}{c} \bf U\\\oOmega\end{array} \right)
= \m
\cdot
\left( \begin{array}{c}
{\bf F}+{\bf F}_0\\
{\bf T}_0
\end{array}\right),\label{main}
\eeq
where  $\m$ is the mobility matrix \cite{Ekiel-Wajnryb-book}. The mobility matrix depends on the instantaneous positions $\bm{r} = (\mathbf{r}_1,...,\mathbf{r}_N)$ of all fiber segments. For a given configuration $\bm{r}$, values of ${\bf F}_0$, ${\bf T}_0$ and $\m$ are determined by the HYDROMULTIPOLE numerical algorithm \cite{MEJ-EW}, 
with the wall effects evaluated by the single-wall superposition \cite{Cichocki2000,Blawzdziewicz-Wajnryb}. The evolution of the fiber is determined by numerically solving the set of coupled differential equations,
\beq
d\bm{r}/dt = \bm{U},
\eeq
where $\bm{U}$ is given by Eq.~\eqref{main}.
An adaptive Runge-Kutta method, implemented in the FORTRAN code is used.

\section{Results}\label{III}
\subsection{Basic information}
In our study, 
fibers are initially aligned with the flow. In our frame of reference (see Fig.~\ref{fig:fiber}), the initial configuration $r_i=(x_i,y_i,z_i)$ of all fiber segments satisfies ${y}_i=0,z_i=z_0, i=1, \ldots, N$. This ensures that at all times $t$, the fiber stays in the plane $y=0$. 
In the performed numerical simulations we fix the flowing values:
\ba
 k&=&10,\\
 l_0&=& 1.05,\\
 h&=&50.
 \ea
A large value of $k$ ensures that the fiber holds an almost constant length of 
\ba
 L &{\approx}&(N-1) l_0+1,
 \label{eq:length}
\ea
with the maximal relative increase from the equilibrium value \eqref{eq:length} of  7\% or less, and the minimal relative decrease not exceeding 4\%. 

In this work, we analyze the dependence of the fiber dynamics on its initial position $z_0$ across the channel, its bending stiffness $A$ and 
the number of beads $N$. The goal is to determine the range of the parameters which corresponds to the fastest fiber migration toward the middle plane of the channel. 

\subsection{Typical evolution of a single fiber}

\begin{figure}[ht]
\includegraphics[width=7.1cm]{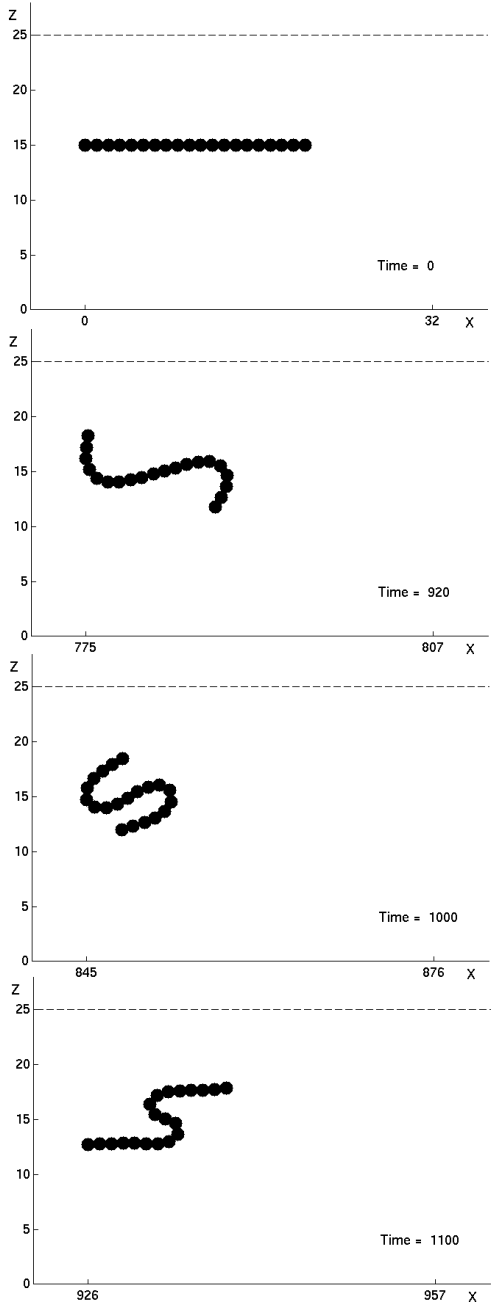}\vspace{-0.4cm}
\caption{Typical evolution of the fiber shape. Number of beads $N=20$, the initial fiber position $z_0=15$ and the bending stiffness $A=0.5$. The tumbling time between the first and the second flipping moment $\tau = 2585$. } \label{figEvo1}
\end{figure}

The motion of a single fiber strand in Poiseuille flow shows a generic behavior. 
Initially aligned with the flow, it slowly begins a tumbling motion, which is almost periodic. Almost, not exactly, because the fiber has a tendency to migrate across the 
flow. 
Fig. \ref{figEvo1} 
shows subsequent snapshots 
from a typical fiber dynamics. 

To quantify evolution of the fiber position and shape,  we define the following quantities:
\begin{itemize}
\item $\tau$ is the tumbling time, defined as the time between the 
two consecutive 
crossing of the fiber end-to-end vector (which links the centers of the first and the last beads) with the plane perpendicular to the flow direction. The tumbling time depends on $A,\;N$ and the fiber position $z$ across the channel.
\item $x_{mid}(t)$ and $z_{mid}(t)$ denote the $x$ and $z$ position components of the middle segment of the fiber, if $N$ is odd, and the average position of the beads with labels $N/2$ and $N/2+1$, if $N$ is even.
\item $x_{M}(t)$and $z_{M}(t)$ denote the $x$ and $z$ components of the fiber center-of-mass position. The center is calculated as the simple arithmetic mean over positions of all fiber segments.
\item $\delta x(t)$ and  $\delta z(t)$ denote the $x$ or $z$ components of the difference between the positions of the last and the first segments of the fiber.
\end{itemize}

\subsection{Dependence on initial position across the channel}
The dependence of fiber dynamics on the initial position $z_0$ across the channel has been analyzed for a fiber with $N=20$ segments and the stiffness $A=0.5$.

\begin{figure}[hb]
\includegraphics[width=8.5cm]{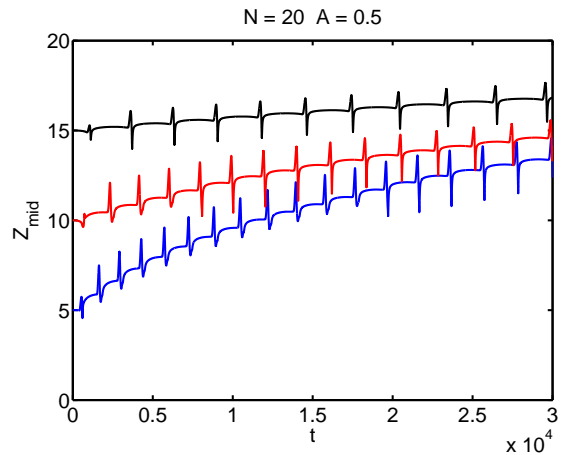}
\caption{Position $z_{mid}$ of the middle section of the fiber as function of time, for different initial positions $z_0$ of the fiber aligned with the flow.} \label{fig2}
\end{figure}

In Fig. \ref{fig2}, the position $z_{mid}$ of the middle point of the fiber is shown as a function of time, for different initial positions $z_0$. 
For $z_0 = 5, 10, 15$ the evolution of fibers is characterized by tumbling and migration towards the center of the channel. 
Both phenomena depend on the instantaneous distance from the wall $z_{mid}(t)$. Fibers placed exactly on the line of symmetry of the channel move along 
with the flow without changing their relative configuration. 

\begin{figure}[t]
\includegraphics[width=8.5cm]{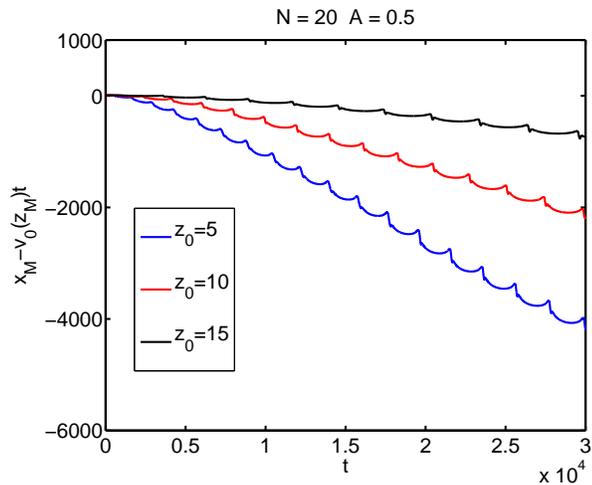}
\caption{Difference between the fiber center-of-mass position $x_{M}(t)$ along the flow and the reference position of the fluid element $\hat{\bm{x}}\cdot \bm{v}_0(z_{M}(t))t$, moving with the Poiseuille flow, for different initial positions $z_0$ of the fiber.} \label{fig1}
\end{figure}

Fig. \ref{fig1} shows the difference 
between the center of mass position $x_M(t)$ of the fiber along the flow, and the position of an element of the fluid moving with  the ambient Poiseuille flow 
calculated at  $z_{M}(t)$. The fiber center-of-mass lags behind the ambient flow \eqref{eq:Poiseuille}. For the considered initial positions, this effect is larger for a smaller value of $z_0$, i.e. for a fiber which migrates faster towards the middle plane of the channel. 

\begin{figure}[b]
\includegraphics[width=9.3cm]{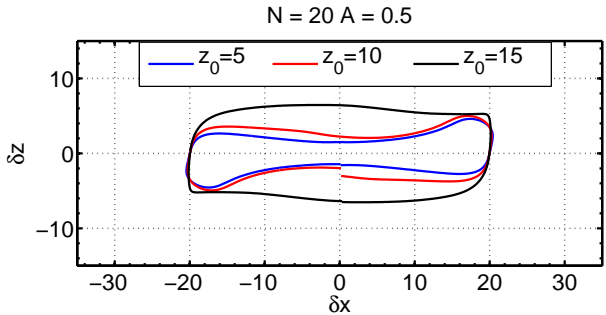}
\caption{Difference $\delta z$ in position of the last and first section of the fiber in the perpendicular to the flow ($z$) direction as a function of the difference $\delta x$ in position of the last and the first sections of 
the fiber in the flow ($x$) direction, for different initial positions $z_0$. }\label{fig4}
\end{figure}

In Fig. \ref{fig4}, $\delta z$ is plotted versus $\delta x$  
for different initial positions $z_0$ of the fiber. Each trajectory corresponds to two tumbling times: it starts from the 2nd flipping moment and ends on the 4th flipping moment of the fiber. 
As expected,  $\delta x|_{\delta z =0}$ practically does not depend on $z_0$ - length of fibers aligned with the flow is insensitive to their position across the channel. However, $\delta z|_{\delta x =0}$ is larger for larger distances from the wall, where fibers deform less.
For a smaller distance from the wall, the fiber shape is more deformed in comparison to the initially straight line. Rigid non-deformable rods would just rotate, and their trajectories would be circles.

Conclusions from this subsection are the following.
\begin{enumerate}
  \item For $z\ge 5$, fibers 
 perform a tumbling motion. The tumbling time increases when the distance from the wall for $z$ becomes larger.
  \item Migration rate of fibers towards the center of the channel is larger for a smaller distance $z$ from the wall (at least for $z\ge 5$).  
  \item Flipping tends to hinder the motion of the fiber along the flow, in comparison to the ambient Poiseuille flow.
  \item The faster fiber migration across the channel, the larger hindrance of its motion along the channel. 
\item For a larger distance from the wall, the shape of a tumbling fiber is less deformed.
\end{enumerate}

\subsection{Dependence on length (number of segments)}
The equilibrium length of a fiber is given in terms of the number of beads $N$ used to model it, and the distance $l_0$ between them, see Eq. \eqref{eq:length}. In this section, we analyze how the fiber dynamics 
depends on the number of beads~$N$. We fix the stiffness parameter $A=0.5$ and the initial position $z_0=10$.

\begin{figure}[ht]
\includegraphics[width=8.3cm]{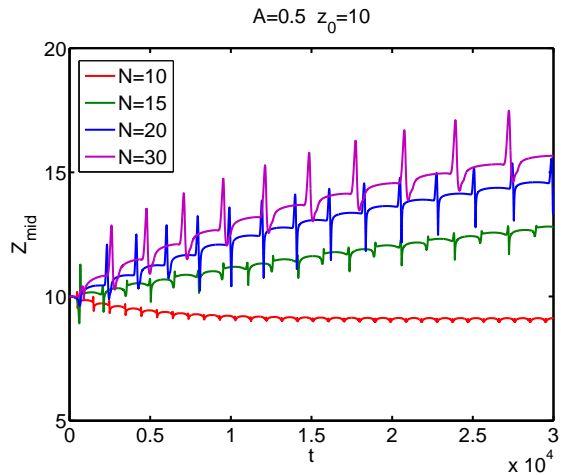}
\caption{Position $z_{mid}$ of the middle section of the fiber as a function of time for different numbers $N$ of segments in the fiber aligned initially with the flow at $z_0 =10$. 
}\label{fig11}
\end{figure}

Fig. \ref{fig11} shows the position $z_{mid}$ as a function of time, for fibers made of a different number of segments $N$. All fibers are initially placed at $z=10$ and aligned with the flow. For a sufficiently large number of beads $N$, we observe that longer fibers 
migrate 
faster toward the middle of the channel, and their tumbling time is slightly longer.  The shortest fiber, with $N=10$, tumbles and 
migrates slowly towards the wall. 

\begin{figure}[t]
\includegraphics[width=8.5cm]{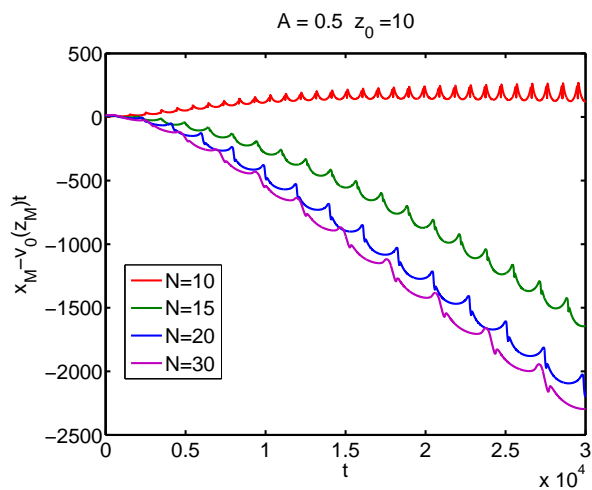}
\caption{Difference between the fiber center-of-mass position $x_{M}(t)$ along the flow, and the reference position $\hat{\bm{x}}\cdot\bm{v}_0(z_{M}(t))t$ of the fluid,  for fibers made of different numbers of beads $N$. 
 }\label{fig10}
\end{figure}

Fig. \ref{fig10} shows the difference between the fiber center-of-mass position along the flow, and the reference position of the fluid element moving with the ambient Poiseuille flow velocity, see Eq.~\eqref{eq:Poiseuille}, 
calculated at  $z_{M}(t)$. Longer fibers move slower than the ambient flow, and the shortest fiber with $N=10$ 
moves faster. The fibers which migrate towards (away from) the wall
are faster (slower) than the ambient Poiseuille flow. In agreement with the observations from the previous subsection, 
hindrance of the motion along the flow is correlated with the enhancement 
of the migration rate towards the middle plane of the channel. 

\begin{figure}[ht]
\includegraphics[width=8.5cm]{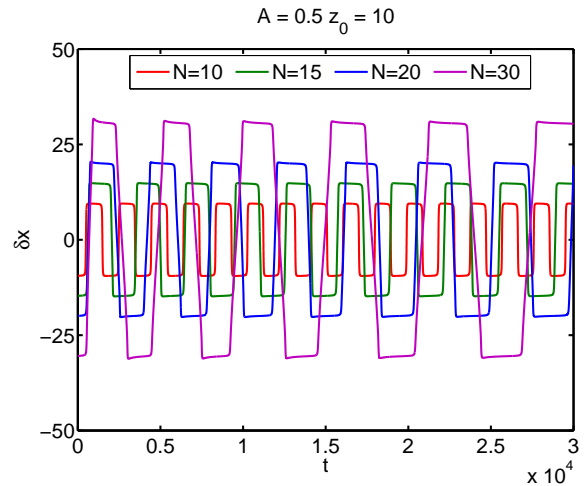}
\caption{Difference $\delta x$ in position  of the last and the first section of the fiber in the flow ($x$) direction as a function of time, for fibers made of different numbers of beads $N$. Initially the fiber is aligned with the flow at $z_0 =10$.}\label{fig12}
\end{figure}

Fig. \ref{fig12} shows $\delta x$ as a function of time,  
for fibers made of different numbers of beads $N$. This plot clearly shows the lengthening of the tumbling time with the increasing length of the fiber. 
Note that for a shorter fiber, the times it takes to flip is much shorter, compared to the time it spends stretched out in the flow. In particular, for $N=10$, the plot of  $\delta x(t)$ consists of almost horizontal sections connected by almost vertical segments.  

\begin{figure}[ht]
\includegraphics[width=9cm]{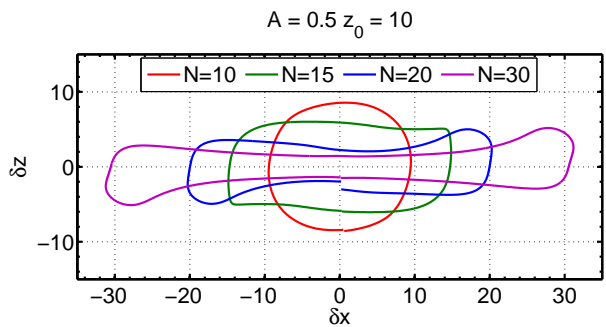}
\caption{Difference $\delta z$ in position of the last and first section of the fiber in the perpendicular to the flow ($z$) direction as a function of the difference $\delta x$ in position of the last and the first sections of 
the fiber in the flow ($x$) direction, for different numbers $N$ of particles in the fiber. Initially the fiber is aligned with the flow at $z_0 =10$. }\label{fig13}
\end{figure}

Fig. \ref{fig13} shows $\delta z$ as a function of $\delta x$, for different numbers of segments in the fiber. The short fiber with $N=10$ segments 
does not fold significantly,
and the plotted evolution curve is almost a circle. The longer the fiber is, the more it bends during tumbling. Note that $\delta z|_{\delta x=0}$ is smaller for longer fibers - the fiber ends are closer to each other at tumbling moments. 

Conclusions from this subsection are the following.
\begin{enumerate}
\item Shorter (longer) fibers move along the flow faster (slower) than the ambient flow, and migrate towards (away from) the wall.
\item Longer fibers migrate faster towards the middle of the channel and move slower along the flow.
\item Longer fibers take much larger fraction of the tumbling time to flip, and bend over much more during that process.
\item Tumbling time increases with the length of the fiber.
\end{enumerate}

\subsection{Dependence on bending stiffness}
The dependence of fiber dynamics on the bending stiffness $A$ has been analyzed for a fiber with $N=20$ segments and the initial position $z_0=5$  across the channel.

\begin{figure}[t]
\includegraphics[width=8cm]{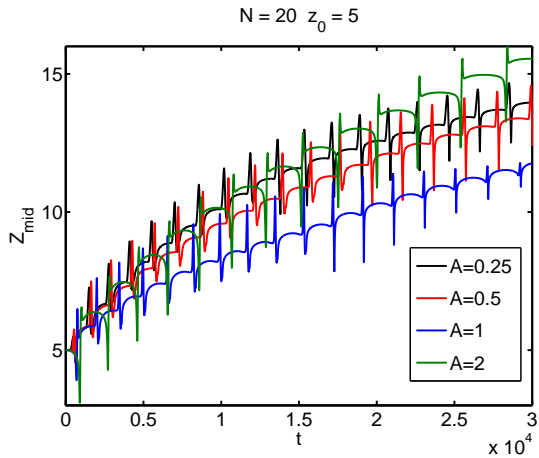}
\caption{Position $z_{mid}$ of the middle section of the fiber as a function of time, for different values of the bending stiffness~$A$. 
}\label{fig6}
\end{figure}

Fig. \ref{fig6} shows the position $z_{mid}$ as a function of time for different values of the bending stiffness $A$. The fibers 
move with tumbling towards the central plane of the 
channel. For the bending stiffness $A\!=\!0.25,0.5$ and~$1$,
migration rate of these fibers across  the channel increases when the stiffness $A$ of the fiber becomes smaller. However, the migration rate for $A=2$ is larger than for $A$=1. 

\begin{figure}[ht]
\includegraphics[width=8cm]{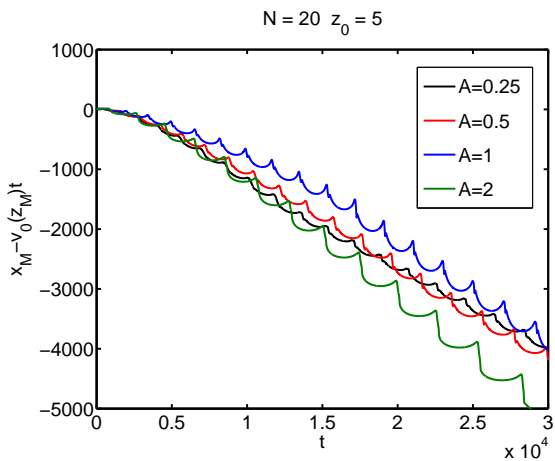}
\caption{Difference between the center of mass position $x_{M}(t)$ of the fiber along the flow and the reference position of the fluid element $\hat{\bm{x}}\cdot\bm{v}_0(z_{M}(t))t$, for different values of the bending stiffness~$A$. }\label{fig7}
\end{figure}

Concurrently, as shown in Fig \ref{fig7}, 
the fibers which migrate towards the middle plane of the channel,
are slower than the ambient Poiseuille flow.
In agreement with the previous observations,
more pronounced hindrance of the motion along the flow is correlated with the enhancement of the migration rate towards the middle plane of the channel. 


\begin{figure}[ht]
\includegraphics[width=9cm]{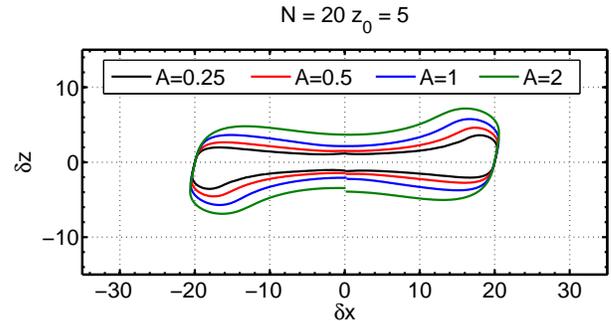}
\caption{Difference $\delta z$ in position of the last and first section of the fiber in the perpendicular to the flow ($z$) direction as a function of the difference $\delta x$ in position of the last and the first sections of 
the fiber in the flow ($x$) direction, for different values of the bending stiffness~$A$.}\label{fig8}
\end{figure}

Fig. \ref{fig8} shows $\delta z$ as a function of $\delta x$ for different flexibilities of fibers at $N=20$ and initial position at $z_0 =5$. As expected, $\delta z|_{\delta x =0}$ is larger for more stiff fibers (which deform less),
while $\delta x|_{\delta z =0}$ does not depend on $A$ (length of fibers aligned with the flow is insensitive to $A$).

Conclusions from this subsection are the following.
\begin{enumerate}
\item Tumbling fibers migrating towards the middle plane of the channel  lag behind the ambient Poiseuille flow.
\item The faster fiber migration toward the middle plane of the channel, the larger hindrance of its motion along the channel. 
\item Larger values of $A$ (greater stiffness) correspond to slightly larger tumbling times.
\item The higher the bending stiffness $A$, the less folded is the fiber when tumbling.
\end{enumerate}

\subsection{Dependence on the magnitude of the Poiseuille flow}
In this section, we analyze the dependence of fiber dynamics on the magnitude of the Poiseuille flow. For the comparison, we have chosen $N=20$ and~$z_0=5$. If we decrease the amplitude $v_m$ of the Poiseuille flow \eqref{eq:Poiseuille} by a factor $\alpha$, then it follows that the relative fiber stiffness $A$ and the relative elastic parameter $k$ (which are inversely proportional to $v_m$) increase by a factor of $\alpha$, see Eqs.~\eqref{kdef} and~\eqref{Adef}. 

Here we take as the reference $A=0.25$ and $k=5$. The decrease of the flow magnitude  $\alpha=2,4,8$ times, corresponds to $(A,k)=(0.5,10),\;(1,20),\;(2,40)$, respectively. In this section, we discuss the resulting change of the fiber dynamics. 

\begin{figure}[ht]
\includegraphics[width=8.5cm]{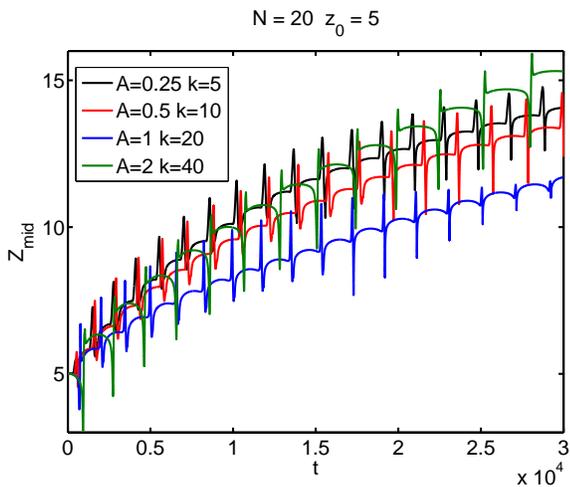}
\caption{Position of the fiber middle-point $z_{mid}$, as a function of time, for decreasing flow amplitudes, i.e. increasing values of ($A,\;k$).
}\label{fig14}
\end{figure}

\begin{figure}[ht]
\includegraphics[width=8.5cm]{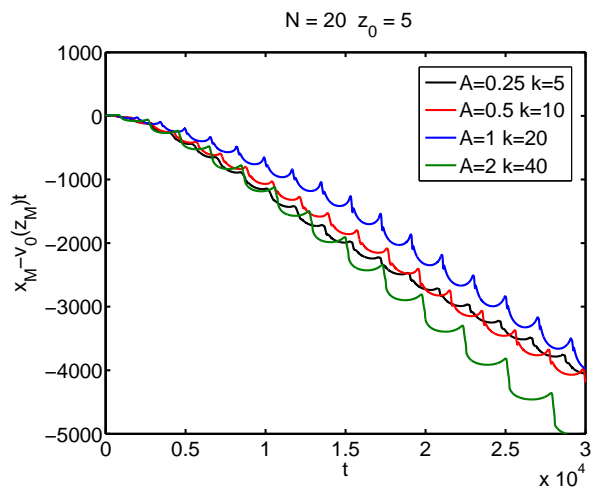}
\caption{Difference between the center of mass position $x_{M}(t)$ of the fiber along the flow and the reference position of the fluid element $\hat{\bm{x}}\cdot\bm{v}_0(z_{M}(t))t$, as a function of time, for decreasing flow amplitudes, i.e. increasing values of ($A,\;k$).
}\label{fig16a}
\end{figure}

Fig. \ref{fig14} shows $z_{mid}$ as a function of time, initial position of the fiber and the flow intensity. The number of segments $N=20$. Comparing the curves for $(A,k)=(0.25,5),\;(0.5,10),\;(1,20)$,  we see that the smaller flow intensity causes slower migration towards the middle of the channel. However, the migration rate is again enhanced for still smaller flows corresponding to $(A,k)=(2,40)$. 
As shown in Fig \ref{fig16a}, 
the fibers which migrate towards the middle plane of the channel,
are slower than the ambient Poiseuille flow.
In agreement with the previous observations,
more pronounced hindrance of the motion along the flow is correlated with the enhancement of the migration rate towards the middle plane of the channel. 

\begin{figure}[ht]
\includegraphics[width=8.5cm]{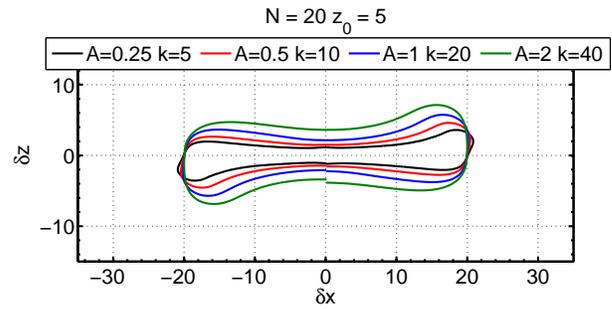}
\caption{Difference $\delta z$ in position  of the last and first section of the fiber in the perpendicular to the flow ($z$) direction as functions of the difference $\delta x$ in position of the last and first section of 
the fiber in the flow ($x$) direction.}\label{fig15}
\end{figure}

An increase of the flow amplitude (i.e. decrease of $A$ and $k$) causes also a decrease of the value of $\delta z|_{\delta x =0}$, and more generally, a larger fiber deformation, as shown in Fig. \ref{fig15}. 


\newpage
Conclusions from this subsection are the following.
\begin{enumerate}
\item Increased flow amplitude leads to a larger fiber deformation;
 fibers tend to keep folded along the flow rather than to stretch across the flow.
\item The faster fiber migration toward the middle plane of the channel, the larger hindrance of its motion along the channel. 
\end{enumerate}

\section{Conclusions}\label{IV}
Dynamics of single flexible fibers entrained by the Poiseuille flow in a wide channel made of two parallel solid plane walls has been evaluated with the use of accurate  numerical code {\sc hydromultipole}. The fibers are initially aligned with the flow, at a distance $z_0$ from the wall. For moderate values of $z_0$, and of the fiber bending stiffness $A$ and  its aspect ratio $N$, 
the following generic behavior has been observed. 

In our computations, fiber deformation increases when they are closer to a wall, or consist of a larger number of beads, or are more flexible, or are entrained by a faster ambient flow. 
Moreover, faster migration rate towards the middle plane of the channel is accompanied by a slower translation along the flow, with velocities smaller than the ambient Poiseuille flow. 

For a wide (but not the whole) range of 
values for $z_0$, $A$ and $N$, 
migration rate towards the channel middle plane increases for fibers, which are closer to a wall, or more easy to bend (less stiff), or longer, or entrained by a faster ambient flow. A more detailed studies are under progress, and will be published elsewhere. 

\acknowledgments
M.L.E.J., E.W. and A.M.S. were supported in part by the Polish Ministry of Science and Higher Education, grant N N501 156538.
M.L.E.-J., E.W. and A.M.S. thank Jerzy B\l awzdziewicz for helpful discussions. 

\end{document}